\documentclass[12pt]{article}
\newcommand{\beq}{\begin{equation}}
\newcommand{\eeq}{\end{equation}}
\newcommand{\beqa}{\begin{eqnarray}}
\newcommand{\eeqa}{\end{eqnarray}}
\newcommand{\beqar}{\begin{eqnarray*}}
\newcommand{\eeqar}{\end{eqnarray*}}

\newcommand{\inn}{\!\cdot\!}

\newcommand{\z}{\zeta}

\newcommand{\eg}{{\it e.g.,}\ }
\newcommand{\ie}{{\it i.e.,}\ }
\newcommand{\labell}[1]{\label{#1}} 
\newcommand{\reef}[1]{(\ref{#1})}
\newcommand\prt{\partial}

\newcommand\tG{{\widetilde G}}

\newcommand\ta{{\tilde a}}
\newcommand\tb{{\tilde b}}
\newcommand\tc{{\tilde c}}
\newcommand\td{{\tilde d}}

\newcommand\ti{{\tilde i}}
\newcommand\tj{{\tilde j}}

\begin{document}

\begin{titlepage}

\begin{center}



\vskip 2 cm
{\LARGE \bf   T-duality of   $\alpha'$-correction to  DBI action  \vskip 0.6 cm   at all orders of gauge field  
 }\\
\vskip 1.25 cm
  Mohammad R. Garousi\footnote{garousi@ um.ac.ir}

\vskip 1 cm
{{\it Department of Physics, Ferdowsi University of Mashhad\\}{\it P.O. Box 1436, Mashhad, Iran}\\}
\vskip .1 cm
\vskip .1 cm

\end{center}

\vskip 0.5 cm

\begin{abstract}
\baselineskip=18pt
By explicit calculations of four-field couplings, we observe that the higher derivative corrections to the DBI action   in flat space-time, can be either in a covariant form or in a T-duality invariant form. The two forms are related by a non-covariant field redefinition.  Using this observation, we then propose a non-covariant but T-duality invariant action   which includes all  orders of massless  fields and has two extra derivatives with respect to the DBI action.  

\end{abstract}
\end{titlepage}
\section{Introduction}
The effective action of a D$_p$-brane in bosonic string theory includes various world-volume couplings of open string tachyon, transverse scalar fields,   gauge field, closed string tachyon, graviton, dilaton and B-field. Because of the  tachyons, the  bosonic string theory and its D$_p$-branes are all  unstable. Assuming the tachyons are frozen at the top of their corresponding tachyon potentials,    the    effective action at the leading order of $\alpha'$  in flat spacetime   is then given by   Dirac-Born-Infeld (DBI) action \cite{Leigh:1989jq,Bachas:1995kx}
\beqa
S_p&\supset&-T_p\int d^{p+1}\sigma\sqrt{-\det(\tG_{ab}+  F_{ab})}\labell{DBI}
\eeqa
 where  $\tG_{ab}$  is   pull-back of the  bulk flat metric onto the world-volume\footnote{Our index convention is that the Greek letters  $(\mu,\nu,\cdots)$ are  the indices of the space-time coordinates, the Latin letters $(a,d,c,\cdots)$ are the world-volume indices and the letters $(i,j,k,\cdots)$ are the normal bundle indices. The killing index in the reduction of 10-dimensional space-time to 9-dimensional space-time is $y$.}, \ie
\beqa
\tG_{ab}&=&P[\eta]_{ab}\,=\,\frac{\prt X^{\mu}}{\prt\sigma^a}\frac{\prt X^{\nu}}{\prt\sigma^b}\eta_{\mu\nu}\nonumber\\
 &=&\eta_{ab}+ \prt_a\chi^{i}\prt_b\chi^{j}\eta_{ij}\labell{pull}
\eeqa
where in the second line the pull-back is in   static gauge\footnote{In the literature, there is a factor of $2\pi\alpha'$ in front of gauge field strength $F_{ab}$ in the DBI action. We normalize the gauge field to  absorb this factor. With this normalization, the gauge field along the killing direction, $A_y$, transforms to  the transverse scalar field $\chi^y$ under T-duality.}, \ie $X^i=\chi^i$ and $X^a=\sigma^a$.   The DBI action \reef{DBI} is covariant under the general coordinate transformations  and is invariant under T-duality\footnote{By invariance under T-duality, we mean after expanding the action to a specific order of field, the couplings at each order  satisfy the T-duality constraint \reef{Tconstraint}.}.  With our normalization for the gauge field, the DBI action is at the leading order of $\alpha'$.  The first correction  to this action   is at order $\alpha'$ in which we are interested in this paper. The $\alpha'$ corrections to Born-Infeld action which includes only gauge field, have been studied in \cite{Abouelsaood:1986gd,Tseytlin:1987ww,Andreev:1988cb, Wyllard:2000qe, Andreev:2001xx, Wyllard:2001ye}. Using the proposed connection between relativistic hydrodynamics and open string effective action \cite{ Niarchos:2015moa,Grignani:2016bpq}, the $\alpha'$ corrections to the DBI action may be used for studying higher-derivative corrections in relativistic hydrodynamics. 

At zero gauge field level,  the general covariance requires the world-volume couplings at any order of $\alpha'$ consist of various contractions of   the second fundamental form $\Omega_{ab}{}^{\mu}=D_a\frac{\prt X^{\mu}}{\prt\sigma^b}$ and its covariant derivatives.   By studying the disk-level S-matrix element of two graviton vertex operators at low energy,   such  couplings at order $\alpha'$  have been found in   \cite{Corley:2001hg}   to be 
\beqa
S_p& \supset&-T_p\int d^{p+1}\sigma\sqrt{-\det(\tG_{ab}   )}\bigg[1+\frac{\alpha'}{2}R+\alpha'\widetilde{\bot}_{\mu\nu} \tG^{ab}\tG^{cd} \bigg(\Omega_{ab} {}^\mu \Omega_{cd}{}^\nu- \Omega_{ac}{}^\mu \Omega_{bd}{}^\nu \bigg) \bigg]\labell{DBI1} 
\eeqa
where $R=\widetilde{G}^{\mu\nu}\widetilde{G}^{\alpha\beta}R_{\mu\alpha\nu\beta}$ and $\widetilde{G}^{\mu\nu}$  is  the first fundamental form, \ie
\beqa
\widetilde{G}^{\mu\nu}&=&\frac{\prt X^{\mu}}{\prt\sigma^a}\frac{\prt X^{\nu}}{\prt\sigma^b}\widetilde{G}^{ab},
\eeqa
which is a projection operator, \ie $\eta_{\nu\alpha}\widetilde{G}^{\mu\nu}\widetilde{G}^{\alpha\beta}=\widetilde{G}^{\mu\beta}$. It projects space-time tensors  to the world-volume space.   The tensor $\widetilde{\bot}_{\mu\nu}$ in \reef{DBI1} is a projection operator, \ie $\eta^{\nu\alpha}\widetilde{\bot}_{\mu\nu}\widetilde{\bot}_{\alpha\beta}=\widetilde{\bot}_{\mu\beta}$, which  projects space-time tensors  to the transverse  space. This projection operator can be written in terms of the first fundamental form as
\beqa
\widetilde{\bot}_{\mu\nu}&=&\eta_{\mu\nu}-\widetilde{G}_{\mu\nu}\labell{tbot}
\eeqa  
In flat spacetime and in the static gauge, the  transverse component of the second fundamental form  in \reef{DBI1}  is 
\beqa
\Omega_{ab}{}^i&=&\prt_a\prt_b\chi^i-\widetilde{\Gamma}_{ab}{}^c\prt_c\chi^i\nonumber\\
&=&\prt_a\prt_b\chi^i-\widetilde{G}^{ij} \prt_a\prt_b\chi^k\eta_{jk} \nonumber\\
&=&\widetilde{\bot}^i{}_{j}\prt_a\prt_b\chi^j\labell{cov1}
\eeqa
 and its world-volume component $\Omega_{ab}{}^c$ is zero.

The   covariant  action \reef{DBI1} includes  all  orders   of the transverse scalar fields. Its two-scalar couplings    do not change the scalar   propagator as expected from S-matrix elements. We have also checked that its four-scalar couplings are   consistent with the disk-level S-matrix element of four gauge bosons in the bosonic string theory which calculated several years ago \cite{Kawai:1985xq} (see Appendix). 

In this paper, we are interested in including the   gauge field $A_a$ into the action \reef{DBI1}. To this end, one may consider all gauge invariant couplings with unknown coefficients and impose the consistency of the couplings with the corresponding S-matrix elements to constrain the coefficients. There are further constraints if one requires the couplings to be covariant under the general coordinate transformations and/or to be invariant under  T-duality transformations. We shall show that unlike the DBI action \reef{DBI}, the higher covariant  derivative corrections to this action is not invariant under T-duality. As a result, we have to constraint the higher derivative couplings  to be consistent with the S-matrix elements and to be either  covariant under the general coordinate transformations   or to be invariant under the T-duality transformations. In the latter case, we propose a T-duality invariant action which includes all orders of the gauge field. 

An outline of the paper is as follows: In the next section, using the consistency of the couplings in flat spacetime with the general coordinate transformations and with the S-matrix element of four gauge bosons, we construct a particular four-field couplings at order $\alpha'$. We then show that the couplings are not consistent with the T-duality transformations. To construct a T-duality invariant action, in section 3, we propose a non-covariant  extension for   the  pull-back metric which is   invariant under T-duality when the   metric is not along  the killing direction. Moreover, we   introduce new tensors which are the transformations of the above object under the T-duality when it is along   the killing direction. Using these non-covariant objects, we then construct a    T-duality invariant action at order $\alpha'$ which includes all orders of the gauge field and is consistent with the S-matrix element of four gauge bosons. In section 4, we discuss our results.

\section{Covariant action}

In this section we are going to find four-field couplings at order $\alpha'$ which are consistent with the S-matrix elements and are covariant under the general coordinate transformation. The latter condition requires the couplings to be constructed from the gauge field strength $F_{ab}$, its covariant derivative, \ie
	\beqa
	D_aF_{bc}&=&\prt_aF_{bc}- \widetilde{\Gamma}_{ab}{}^dF_{dc}-\widetilde{\Gamma}_{ac}{}^dF_{bd}\nonumber\\
	&=&\prt_aF_{bc}-\widetilde{G}^d{}_i\prt_a\prt_b\chi^iF_{dc}+\widetilde{G}^d{}_i\prt_a\prt_c\chi^iF_{db}\labell{cov2}
	\eeqa
and the second fundamental form $\Omega_{ab}{}^i$. 

The couplings of two transverse scalars and two gauge fields may have structure $FF\Omega\Omega$ or $DFDF$. We extend the square root in \reef{DBI1} to the DBI action, \ie
\beqa
\sqrt{-\det(\tG_{ab}   )}&\rightarrow &\sqrt{-\det(\tG_{ab}+F_{ab}   )}
\eeqa
Upon expand the square root, one   finds a  particular set of couplings with structure  $FF\Omega\Omega$. However, we have checked that they are not consistent with corresponding S-matrix element, \ie they are not consistent with \reef{2A2X}. As a result, the couplings may have other terms with structure  $FF\Omega\Omega$   and/or couplings with structure $DFDF$.
   The latter couplings, however,   should not change the propagator of gauge fields. So they must be total derivative  at two-field level, \ie
	\beqa
S_p& \supset&	-T_p\int d^{p+1}\sigma\sqrt{-\det(\tG_{ab}+F_{ab} )}\bigg[1+\alpha'\left(\frac{\alpha}{2}D_aF_{bc}D^aF^{bc}+\alpha D_aF^a{}_{b}D_cF^{bc} \right) \bigg]\labell{DFDF}
	\eeqa
where the world-volume indices   are raised by $\widetilde{G}^{ab}$ and $\alpha$ is a constant.    There are also 6 new couplings with structure  $FF\Omega\Omega$.   The consistency with the S-matrix element, \ie \reef{2A2X}, can not uniquely fix the coefficients of these couplings. For example, if we do not consider the  new couplings with structures $FF\Omega\Omega$, then the S-matrix  fixes $\alpha=1$.	 

To study  four gauge field couplings at order $\alpha'$, one may consider  all     couplings with structure $FFDFDF$   with unknown coefficients and  constrains   the coefficients by comparing them with the corresponding S-matrix element, \ie comparing with \reef{4A}. There are 18 such couplings and the comparison with \reef{4A} produces only three constraints. So  three couplings are fixed and all others have   unconstrained coefficients. One particular choice for the coefficients are 
\beqa
S_p& \supset&	-T_p\int d^{p+1}\sigma\sqrt{-\det(\tG_{ab}+F_{ab} )}\bigg[1+\alpha'\left(\frac{1}{6} F_a{}^e F_{de} D^aF^{bc} D^dF_{ bc}\right.\labell{FFDFDF}\\&&\left.\qquad\qquad\qquad\qquad\qquad\qquad\quad
-\frac{1}{3} F_c{}^e F_{de} D^aF^{bc} D_aF_b{}^d -\frac{1}{6} F_{de} F^{de} D_aF_{ bc} D^aF^{bc}\right) \bigg]\nonumber
\eeqa
  The above  couplings  have been found   in  \cite{Andreev:1988cb} by other means. 
	
We have seen that there are many covariant couplings in \reef{DFDF}	and \reef{FFDFDF} which are not fixed by the S-matrix element of four gauge bosons at order $\alpha'$. Moreover, it is difficult to continue  the above construction to find six-field couplings and higher orders. So one may use other consistency condition to fix the unknown coefficients at four-field level and possibly for higher orders. We are going to impose the constraint that like the DBI action \reef{DBI}, all higher derivative couplings to be invariant under the T-duality. However,   the T-duality constraint is not consistent with the general covariance constraint when the higher derivative terms are included into the D-brane action. To see this point, we note that under T-duality $F_{ab}$ transforms to $\prt_a\chi^i$   when the gauge field is along the killing direction. On the other hand, while $F_{ab}$ appears at various places in \reef{FFDFDF}, the velocity  $\prt_a\chi^i$ appears only in the  definition of the  covariant derivative \reef{cov2} and in the pull-back metric \reef{pull}. 
	 In fact, we have considered all possible covariant couplings with unknown coefficients and constrained them to satisfy the T-duality invariance of the corresponding action. This produces a set of constraints which are not consistent with the constraints from the S-matrix elements. Therefore, the D-brane effective action at order $O(\alpha')$   which is consistent with the S-matrix elements and is invariant under T-duality may not be covariant under general coordinate transformations. In the next section we study the effective action which is invariant under T-duality transformation.

\section{T-duality invariant action}

 The low energy expansion of an arbitrary S-matrix element can be separated into two parts. One part includes massless poles and the other part includes contact terms, \ie
\beqa
{\cal A}&=&{\cal A}_{\rm pole}+{\cal A}_{\rm contact}
\eeqa
The D-brane effective action must reproduce the contract terms in ${\cal A}_{\rm contact}$ in the momentum space. On the other hand, it is known that the S-matrix element of an arbitrary number of gauge or transverse scalar vertex operators must satisfy the Ward identity corresponding to the T-duality \cite{Garousi:2011we,Velni:2012sv}. Since the T-duality is a global duality, ${\cal A}_{\rm pole}$ and ${\cal A}_{\rm contact}$ each must satisfy the Ward identity. In other words, the effective action must satisfy the linear T-duality after imposing the on-shell relations. Since the T-duality  transformations on   gauge   and the scalar fields are linear, \ie $A_y\leftrightarrow \chi^y$, the effective action must be consistent with an action which is   invariant under the T-duality transformations. As we have observed in the previous section, the effective action, however, may not be in a covariant form under the general coordinate transformations.

 Therefore, one should  consider the  non-covariant form for the couplings, \ie
\beqa
\prt F\prt F+ FF\prt F\prt F+ FFFF\prt F\prt F+ FFFFFF\prt F\prt F+\cdots\labell{ser1}
\eeqa
where dots represent higher order terms and the terms containing $\prt\chi$, $\prt\prt\chi$. The indices   are contracted with flat metric\footnote{One may also consider terms which contain the second  derivative of gauge field strength and the third derivative of scalar fields, \ie 
\beqa
  FFF\prt\prt F+ FFFFF\prt\prt F+ FFFFFFF\prt\prt F+\cdots
\eeqa
However, such terms convert to the couplings in \reef{ser1} by integration by part. So in the action we do not consider such couplings.}. Each term has many different contractions. One should contribute an unknown constant for the coefficient of each contraction. One should then impose the condition that the action must be invariant under the T-duality. This constrains the coefficients. The resulting T-dual action must be also consistent with the S-matrix elements. This produces some further constraints on the coefficients. 

Imposing the above two conditions, in principle, one can find four-field couplings, six-field couplings and so on. Since the T-duality is linear, there is no relation between these sets of T-dual couplings, \eg the T-duality produces  no relation between   four-field couplings and six-field couplings.  To find a relation between the above T-dual sets, we use an extra condition that the couplings should be consistent with all orders of  scalar couplings in \reef{DBI1}.   To implement this latter condition, we propose to add appropriate gauge field   into the pull-back metric and into the projection operator $\widetilde{\bot}_{\mu\nu}$ to make the couplings in \reef{DBI1} to be consistent with the T-duality.  We begin by considering an extension for  the pull-back metric to be consistent with T-duality. 
	
	When   derivative of the gauge field strength are zero, the T-duality   may  fix the presence of $F_{ab}$  in the inverse of the pull-back of the flat metric  by extending it to the following expression:
\beqa
\tG^{ab}&\longrightarrow& G^{ab}= \left(\frac{1}{P[\eta]+F} P[\eta]\frac{1}{P[\eta]-F}\right)^{ab} \labell{ext}
\eeqa
which is a symmetric matrix. To show that the above replacement   is consistent with T-duality, suppose the D-brane is along the circle on which the T-duality is imposed. When $a,b$ are not the killing index, \ie $a=\ta,b=\tb$, using the prescription given in \cite{Garousi:2009dj}, one can   verify that  $G^{\ta\tb}$ is invariant under T-duality. To the second order, it is
\beqa
G^{\ta\tb}&=&\eta^{\ta\tb}-\prt^{\ta}\chi^i\prt^{\tb}\chi^j\eta_{ij}+F^{\ta c}F ^{ d\tb}\eta_{cd}\nonumber\\
&=&\eta^{\ta\tb}-\prt^{\ta}\chi^i\prt^{\tb}\chi^j\eta_{ij}+F^{\ta\tc}F ^{\td\tb}\eta_{\tc\td}-\prt^{\ta}A_y\prt^{\tb}A_y\eta^{yy}\nonumber\\
&\stackrel{T}{\longrightarrow}&\eta^{ab}-\prt^a\chi^{\ti}\prt^b\chi^{\tj}\eta_{\ti\tj}+F^{ac}F^{db}\eta_{cd}-\prt^a\chi^y\prt^b\chi^y\eta_{yy}\nonumber\\
&=&\eta^{ab}-\prt^a\chi^i\prt^b\chi^j\eta_{ij}+F^{ac}F^{db}\eta_{cd}\,=\,G^{ab}
\eeqa
 We have checked it with the Mathematica to the tenth order and found that $G^{\ta\tb}$  is invariant.  

Inverse of the new metric \reef{ext} is 
\beqa
G_{ab}&=&P[\eta]_{ab}-F_{ac}F_{db}\widetilde{G}^{cd}
\eeqa
which is not the pull-back of the flat metric in the static gauge, so the   extension \reef{ext} does not have geometrical interpretation.   As a result, an action which uses  this matrix is not in a covariant form anymore.   The matrix \reef{ext} defines a new   form with space-time indices, \ie
\beqa
G^{\mu\nu}&=&\frac{\prt X^{\mu}}{\prt\sigma^a}\frac{\prt X^{\nu}}{\prt\sigma^b}G^{ab}\labell{G1}
\eeqa
We will find that all components of this form, \ie $G^{ab}$, $G^{ai}$ and $G^{ij}$ appear in the T-duality invariant    action at order $\alpha'$. Note that $G^{\mu\nu}$ is not projection operator, so it is not a first fundamental form. This matrix has even number of gauge field strength.

When one or both indices of $G^{ab}$ are the killing index, it turns out that it transforms to the following expressions under T-duality:
\beqa
G^{y\ta}&\stackrel{T}{\longrightarrow}&\Theta^{\ta}{}^{ y}\nonumber\\
G^{yy}&\stackrel{T}{\longrightarrow}&\bot^{yy}\labell{Lyy}
\eeqa
Note that the $y$ index on the left hand side is a world-volume index whereas  on the right hand side it is a transverse index. The new matrix $\bot^{ij}=\eta^{ij}-G^{ij}$ and matrix $\Theta^{a}{}^{i}$ is the following:
\beqa
\Theta^{a}{}^{i}&=&\left(\frac{1}{P[\eta]+F} F\frac{1}{P[\eta]-F}\right)^{ab}\prt_b\chi^i 
\eeqa
We have checked the T-duality transformations \reef{Lyy}   to the tenth order  and found agreement. The presence of $\Theta^{a}{}^{i}$ indicates that in a T-duality invariant theory there must be another tensor as
\beqa
\Theta^{\mu\nu}&=&\frac{\prt X^{\mu}}{\prt\sigma^a}\frac{\prt X^{\nu}}{\prt\sigma^b}\Theta^{ab}\labell{G2}
\eeqa
where the new antisymmetric matrix $\Theta^{ab}$ is
\beqa
\Theta^{ab} &=&\left(\frac{1}{P[\eta]+F} F\frac{1}{P[\eta]-F}\right)^{ab} \labell{theta}
\eeqa
Note that $G^{ab}-\Theta^{ab}$ is the inverse of the operator inside the square root in the DBI action \reef{DBI}, \ie $\left(\frac{1}{P[\eta]+F}\right)^{ab}=G^{ab}-\Theta^{ab}$. We will find that all components of \reef{G2}, \ie $\Theta^{ab}$, $\Theta^{ai}$  and $\Theta^{ij}$ appear in the effective action at order $\alpha'$. Note that $\Theta^{\mu\nu}$ is not a projection operator. It has odd number of gauge field strength.

We have  also found that $\Theta^{ab}$ when $a,b$ are not the killing index, is invariant under T-duality and found the following T-duality transformations: 
\beqa
\Theta^{y}{}^{i}&\stackrel{T}{\longrightarrow}&-{G}^{iy}\labell{{L_1}y}\\
\Theta^{\ta y} &\stackrel{T}{\longrightarrow}&G^\ta{}^y\nonumber\\
G^y{}^i&\stackrel{T}{\longrightarrow}&\Theta^{iy}\nonumber
\eeqa
which can be verified by expanding the matrices in both sides in terms of $F$ and $\prt\chi$ and using the T-duality transformation $A_y\stackrel{T}{\longrightarrow}\chi^y$. 

 Therefore, in a manifestly  T-duality invariant action there may be the matrices  $G^{\mu\nu}, \Theta^{\mu\nu}$  and $\bot^{ij}$ which form a T-dual multiplet.   Moreover, the action at order $\alpha'$ must include $\prt\prt\chi$ and $\prt F$ which form another multiplet.     We are going to extend the scalar couplings in \reef{DBI1} to include the  gauge field.   This condition constrains the six indices of $(\prt\prt\chi)^2$, $\prt\prt\chi \prt F$ and $(\prt F)^2$ to be contracted with six indices of $GGG$, $GG\bot$, $GG\Theta$ and so on. Taking the symmetries of these tensors, there are 41 different contractions.

Using the Mathematica package ``xAct'' \cite{CS},  one can write all above contractions as
\beqa
\!\!\!\!\!\!\!\!\!\!\!\!\!\!\!&&c_1 G^ {ad}\Theta^ {be}\Theta^ {cf} \psi_{abc} \psi_ {def} + 
  c_2 G^ {ad}\Theta^ {bc}\Theta^ {ef} \psi_  
{abc} \psi_ {def} + 
  c_3 G^ {cd}\Theta^ {ae}\Theta^ {bf} \psi_  
{abc} \psi_ {def} + 
  c_4 G^ {cd}\Theta^ {ab}\Theta^ {ef} \psi_  
{abc} \psi_ {def}\nonumber\\ \!\!\!\!\!\!\!\!\!\!\!\!\!\!\!&& + 
  c_5 G^ {cf}\Theta^ {ae}\Theta^ {bd} \psi_  
{abc} \psi_ {def} + 
  c_6 G^ {cf}\Theta^ {ad}\Theta^ {be} \psi_  
{abc} \psi_ {def} + 
  c_7 G^ {cf}\Theta^ {ab}\Theta^ {de} \psi_  
{abc} \psi_ {def} + 
  c_8 G^ {ae} G^ {bd} G^ {cf} \psi_ {abc} \psi_ {def} \nonumber\\ \!\!\!\!\!\!\!\!\!\!\!\!\!\!\!&&+ 
  c_9 G^ {ad} G^ {be} G^ {cf} \psi_ {abc} \psi_ {def} + 
  c_{10} G^ {df}\Theta^ {ae}\Theta^ {bc} \psi_  
{abc} \psi_ {def} + 
  c_{11} G^ {df}\Theta^ {ab}\Theta^ {ce} \psi_  
{abc} \psi_ {def} + 
  c_{12} G^ {ab} G^ {ce} G^ {df} \psi_ {abc} \psi_ {def}\nonumber\\ \!\!\!\!\!\!\!\!\!\!\!\!\!\!\!&& + 
  c_{13} G^ {bd}\Theta^ {ac}\Theta^ {ij} \omega_ 
{abi} \omega_ {cdj} + 
  c_{14} G^ {ij}\Theta^ {ac}\Theta^ {bd} \omega_  
{abi} \omega_ {cdj} + 
  c_{15} G^ {ac} G^ {bd} G^ {ij} \omega_ {abi} \omega_ {cdj} + 
  c_{16} G^ {ab} G^ {cd} G^ {ij} \omega_ {abi} \omega_ {cdj} \nonumber\\ \!\!\!\!\!\!\!\!\!\!\!\!\!\!\!&&+ 
  c_{17} \bot^ {ij}\Theta^ {ac}\Theta^ {bd} \omega_  
{abi} \omega_ {cdj} + 
  c_{18} G^ {ac} G^ {bd} \bot^ {ij} \omega_ {abi} \omega_ {cdj} + 
  c_{19} G^ {ab} G^ {cd} \bot^ {ij} \omega_ {abi} \omega_ {cdj} + 
  c_{20}\Theta^ {ai}\Theta^ {be}\Theta^  
{cd} \omega_ {aei} \psi_ {bcd} \nonumber\\ \!\!\!\!\!\!\!\!\!\!\!\!\!\!\!&&+ 
  c_{21}\Theta^ {ai}\Theta^ {bd}\Theta^  
{ce} \omega_ {dei} \psi_ {bac} + 
  c_{22}\Theta^ {ai}\Theta^ {bc}\Theta^  
{de} \omega_ {aei} \psi_ {bcd} + 
  c_{23} G^ {bd} G^ {ce}\Theta^ {ai} \omega_ {dei}  
\psi_ {bac} + 
  c_{24} G^ {bc} G^ {de}\Theta^ {ai} \omega_ {aei}  
\psi_ {bcd} \nonumber\\\!\!\!\!\!\!\!\!\!&&+ 
  c_{25} G^ {bc} G^ {de}\Theta^ {ai} \omega_ {dei}  
\psi_ {bac} + 
  c_{26} G^ {cd}\Theta^ {ai}\Theta^ {bj} \omega_  
{acj} \omega_ {bdi} + 
  c_{27} G^ {cd}\Theta^ {ai}\Theta^ {bj} \omega_  
{aci} \omega_ {bdj} + 
  c_{28} G^ {cd}\Theta^ {ai}\Theta^ {bj} \omega_  
{abi} \omega_ {cdj} \nonumber\\ \!\!\!\!\!\!\!\!\!\!\!\!\!\!\!&&+ 
  c_{29} G^ {ai} G^ {bd}\Theta^ {ce} \omega_ {aei}  
\psi_ {bcd} + 
  c_{30} G^ {ai} G^ {be}\Theta^ {cd} \omega_ {aei}  
\psi_ {bcd} + 
  c_{31} G^ {ai} G^ {be}\Theta^ {cd} \omega_ {dei}  
\psi_ {bac} + 
  c_{32} G^ {ai} G^ {ce}\Theta^ {bd} \omega_ {dei} 
\psi_ {abc} \nonumber\\ \!\!\!\!\!\!\!\!\!\!\!\!\!\!\!&&+ 
  c_{33} G^ {ai} G^ {ce}\Theta^ {bd} \omega_ {dei}  
\psi_ {bac} + 
  c_{34} G^ {ai} G^ {de}\Theta^ {bc} \omega_ {aei}  
\psi_ {bcd} + 
  c_{35} G^ {ai} G^ {de}\Theta^ {bc} \omega_ {dei}  
\psi_ {abc} + 
  c_{36} G^ {ai} G^ {de}\Theta^ {bc} \omega_ {dei}  
\psi_ {bac} \nonumber\\ \!\!\!\!\!\!\!\!\!\!\!\!\!\!\!&&+ 
  c_{37} G^ {bj}\Theta^ {ai}\Theta^ {cd} \omega_  
{acj} \omega_ {bdi} + 
  c_{38} G^ {bj}\Theta^ {ai}\Theta^ {cd} \omega_  
{aci} \omega_ {bdj} + 
  c_{39} G^ {ai} G^ {bj} G^ {cd} \omega_ {acj} \omega_ {bdi} + 
  c_{40} G^ {ai} G^ {bj} G^ {cd} \omega_ {aci} \omega_ {bdj} \nonumber\\ \!\!\!\!\!\!\!\!\!\!\!\!\!\!\!&&+ 
  c_{41} G^ {ai} G^ {bj} G^ {cd} \omega_ {abi} \omega_ {cdj}= {\cal{L}}_p \labell{Lp}
	\eeqa 
where   $c_1,\cdots, c_{41}$ are unknown coefficients, 
\beqa
\psi_{abc}\equiv\prt_aF_{bc},&&\omega_{abi}\equiv\prt_a\prt_b\chi_i,
\eeqa
 and the contraction of indices in \reef{Lp} are   with the flat metric. Each term has even number of gauge field which is required by the invariance of the couplings under the parity. When expanding the matrices  $G^{\mu\nu}, \Theta^{\mu\nu}$  and $\bot^{ij}$ around the flat metric, one finds a particular form for the series \reef{ser1} which has only 41 unknown coefficients.  The action which must be invariant under T-duality is  
\beqa
S_p& =&-T_p\int d^{p+1}\sigma\sqrt{-\det(\tG_{ab} +F_{ab} )}\bigg[1+  {\cal{L}}_p +O(\alpha')\bigg]\labell{Sp} 
\eeqa
 where ${\cal{L}}_p$  is given in \reef{Lp}. Some of the couplings in \reef{Sp} are total derivative terms that must be ignored and some other couplings are related to each other by the Bianchi identity. After constraining \reef{Sp} to satisfy the T-duality condition, we will remove the total derivative terms and impose the Bianchi identity to remove the   couplings which are related by the Bianchi identity.

To constraint the   couplings \reef{Sp} to be consistent with the T-duality, following \cite{Garousi:2009dj}, we   reduce the 10-dimensional space-time to the 9-dimensional space-time. It reduces \reef{Sp}  to  two  different actions $S_p^w$ and  $S_p^t$. In $S_p^w$,  the killing direction $y$ is a world-volume direction, \ie $a=(\tilde{a},y)$  and in $S_p^t$   the killing direction   is a transverse direction, $i=(\tilde{i},y)$. The transformation of  $S_p^w$ under the   T-duality   which is called $S_{p-1}^{wT}$, may be equal to $S_{p-1}^t$ up to some total derivative terms which must be ignored in the action, \ie
\beqa
S_{p-1}^{wT}-S_{p-1}^t&=&0\labell{Tconstraint}
\eeqa
This constrains the unknown coefficients in the  Lagrangian \reef{Lp}. Note that if one does not ignore the total derivative terms, then one would find some     constraints which make the total derivative terms in the D-brane action  to be T-duality invariant.

 To impose the above constraint, one may expand  the Lagrangian \reef{Lp} around the flat metric. It is easy to  verify that each term in \reef{Lp} has contribution at the sixth order. So the   T-duality rule \reef{Tconstraint} at order six, constrains all 41 coefficients.   We expect there would be no further constraint  at the eighth order and higher, because the higher order terms are resulted from expanding the matrices  $G^{\mu\nu}, \Theta^{\mu\nu}$  and $\bot^{ij}$ which are already shown to be connected under T-duality transformation to all orders. 

We have expanded the Lagrangian \reef{Lp} to the sixth order   and imposed the   T-duality constraint \reef{Tconstraint}. We have found the following 23 relations between the constants:
\beqa
 &&c_{ 14}\to 0,c_{ 15}\to 0,c_{ 16}\to 0,c_{ 22}\to -2 c_{ 10}+c_{ 11}+2 c_{ 20},c_{ 23}\to -c_{ 12}-2 c_{ 18}-c_{ 19},\nonumber\\&&c_{ 24}\to 2 c_{ 12}-2 c_{ 17}-c_{ 21},c_{ 25}\to c_{ 12}-c_{ 19},c_{ 26}\to -c_{ 18},c_{ 27}\to -c_{ 12}+2 c_{ 17}+c_{ 21},\nonumber\\&&c_{ 28}\to c_{ 12}-2 c_{ 17}-c_{ 19}-c_{ 21},c_{ 29}\to -2 c_{ 10}+c_{ 11}+c_{ 12}+c_{ 19},\nonumber\\&&c_{ 32}\to -3 c_{ 12}-2 c_{ 13}+6 c_{ 17}-3 c_{ 19}+4 c_{ 21}+c_{ 31},c_{ 33}\to c_{ 12}+2 c_{ 13}-2 c_{ 17}+c_{ 19}-c_{ 21}-c_{ 31},\nonumber\\&&c_{ 36}\to -2 c_{ 10}+c_{ 11}-2 c_{ 35},c_{ 37}\to 2 c_{ 12}-2 c_{ 17}+2 c_{ 19}-2 c_{ 21},c_{ 38}\to -2 c_{ 10}+c_{ 11}+c_{ 12}+c_{ 19},\nonumber\\&&c_{ 39}\to -2 c_{ 12}-c_{ 13}+3 c_{ 17}-2 c_{ 19}+2 c_{ 21},c_{ 40}\to c_{ 30}-\frac{c_{ 34}}{2},c_{ 41}\to -2 c_{ 10}+c_{ 11}+c_{ 12}+c_{ 19},\nonumber\\&&c_{ 5}\to 2 c_{ 1}+c_{ 12}+c_{ 13}-3 c_{ 17}+c_{ 19}-2 c_{ 21}-c_{ 3},c_{ 6}\to -2 c_{ 1}-c_{ 12}-c_{ 13}+2 c_{ 17}-c_{ 19}+c_{ 21}+c_{ 3},\nonumber\\&&c_{ 7}\to -4 c_{ 2}+c_{ 30}-\frac{c_{ 34}}{2}+2 c_{ 4},c_{ 9}\to \frac{c_{ 12}}{2}+\frac{c_{ 18}}{2}+\frac{c_{ 19}}{2}-\frac{c_{ 8}}{2}  \labell{rel1}
\eeqa
After inserting  the above relations on the coefficients in \reef{Lp},   one finds 18  T-dual  multiplets. However, not all the coefficients of these multiplets are independent.

We have to impose on these 18 multiplets the condition that the couplings   must be consistent with the scalar couplings in  \reef{DBI1}. This gives the following 5 constraints:
\beqa
 c_{ 12}\to -1+2 c_{ 10}-c_{ 11},c_{ 18}\to -1,c_{ 19}\to 1,c_{ 21}\to 2 c_{ 10}-c_{ 11}+\frac{c_{ 13}}{2}-\frac{3 c_{ 17}}{2},c_{ 34}\to 2 c_{ 30}. \labell{rel2}
\eeqa
We have to also impose the condition that the couplings must be  consistent with the S-matrix element of four gauge bosons at low energy. We have already shown that the four-scalar couplings in \reef{DBI1} are consistent with the corresponding  terms in the  S-matrix element, \ie \reef{4X}. So the above constraints include the consistency of the couplings with the S-matrix element of four scalars.  The consistency of the couplings of four gauge fields with the corresponding terms in the string theory S-matrix element, \ie \reef{4A}, produces the following constraint:  
\beqa
c_{17}\to  \frac{4}{3} + c_{13} \labell{rel3}
\eeqa
To find the above relation, we simply transformed the couplings to the momentum space and impose the on-shell relations and them compare them with the on-shell couplings in \reef{4A}. With the above constraint, the couplings are also consistent with the S-matrix element of two gauge fields and two scalars.

So there are 13 T-dual multiplets in the action which are consistent with \reef{DBI1} and with the S-matrix element of four gauge bosons. There are 12 unknown coefficients in these 13 multiplets. These 12 coefficients   are not   independent yet. We now remove the couplings  which are total derivatives and impose the Bianchi identity in the action. Writing   the gauge field strength in $\prt_a F_{bc}$  in terms of gauge field, \ie$\prt_a\prt_b A_c-\prt_a\prt_c A_b$, one finds the constants $c_1$, $c_2$,$c_3$,$c_4$, $c_8$,$c_{10}$,$c_{20}$,$c_{30}$,$c_{31}$,$c_{35}$ do not appear in the action. That means the   T-dual multiplets   with these coefficients are zero by using the  Bianchi identity. So it is safe to set all these 10 coefficients to zero. There are only two unknown coefficients $c_{11}$ and  $c_{13}$ which appear in the following terms:
\beqa
 &&-T_pc_{11}\int d^{p+1}\sigma\sqrt{-\det(\tG^{ab} +F_{ab} )}\bigg[-\frac{1}{2} G^a{}^d G^b{}^e G^c{}^f \psi_a{}_b{}_c \psi_d{}_e{}_f-G^a{}^b G^c{}^e G^d{}^f \psi_a{}_b{}_c \psi_d{}_e{}_f\nonumber\\&&\qquad\qquad\qquad\qquad\qquad+G^c{}^f \Theta^a{}^e \Theta^b{}^d \psi_a{}_b{}_c \psi_d{}_e{}_f+G^d{}^f \Theta^a{}^b \Theta^c{}^e \psi_a{}_b{}_c \psi_d{}_e{}_f-G^b{}^c G^d{}^e \Theta^a{}^i \psi_b{}_c{}_d \omega_a{}_e{}_i\nonumber\\&&\qquad\qquad\qquad\qquad\qquad+\Theta^a{}^i \Theta^b{}^c \Theta^d{}^e \psi_b{}_c{}_d \omega_a{}_e{}_i-G^a{}^i G^c{}^e \Theta^b{}^d \psi_a{}_b{}_c \omega_d{}_e{}_i+G^b{}^d G^c{}^e \Theta^a{}^i \psi_b{}_a{}_c \omega_d{}_e{}_i\nonumber\\&&\qquad\qquad\qquad\qquad\qquad-G^b{}^c G^d{}^e \Theta^a{}^i \psi_b{}_a{}_c \omega_d{}_e{}_i+G^a{}^i G^d{}^e \Theta^b{}^c \psi_b{}_a{}_c \omega_d{}_e{}_i-\Theta^a{}^i \Theta^b{}^d \Theta^c{}^e \psi_b{}_a{}_c \omega_d{}_e{}_i\bigg]\nonumber
\eeqa
\beqa
 &&-T_pc_{13}\int d^{p+1}\sigma\sqrt{-\det(\tG^{ab} +F_{ab} )}\bigg[ -G^ {bc} G^ {de} \Theta^ {ai}\psi_ {bcd} \omega_ {aei} +
  G^ {cd} \Theta^ {ai} \Theta^ {bj} \omega_ {aci} 
\omega_ {bdj} \nonumber\\&&\qquad\qquad\qquad\qquad\qquad- 
 G^ {cd} \Theta^ {ai} \Theta^ {bj} \omega_ {abi} 
\omega_ {cdj} + 
\bot^{ij} \Theta^ {ac} \Theta^ {bd} \omega_ 
{abi} \omega_ {cdj} + 
 G^ {bd} \Theta^ {ac} \Theta^ {ij} \omega_ {abi} 
\omega_ {cdj}\nonumber\\&&\qquad\qquad\qquad\qquad\qquad + 
 G^ {ai} G^ {ce} \Theta^ {bd}\psi_ {bac} \omega_ {dei} 
- \Theta^ {ai} \Theta^ {bd} \Theta^ {ce} 
\psi_ {bac} \omega_ {dei}\bigg]\labell{total}
\eeqa
The $c_{11}$-terms are zero when gauge field is zero, whereas, the $c_{13}$-terms are zero when the scalar field is zero. We have expanded them to the sixth order and found that they are  total derivative terms. So $c_{11}$ and $c_{13}$ can also be set to zero. 

The final result which is invariant under T-duality, is consistent with the S-matrix element of four gauge bosons and reduces to \reef{DBI1} when gauge field is zero, is the following   action: 
\beqa
S_p&\!\!\!\! \supset\!\!\!\!&	-T_p\int d^{p+1}\sigma\sqrt{-\det(\tG^{ab}+F_{ab} )}\bigg[1+\alpha'\bigg( 
 G^ {ab} G^ {cd}  \bot^{ij}  \omega_ {abi} \omega_ {cdj}- 
 G^ {ac} G^ {bd}  \bot^{ij}  \omega_ {abi} \omega_ {cdj}  \nonumber\\&&\qquad\qquad\qquad\quad
-\frac{1}{2} G^ {ad} G^ {be} G^ {cf}\psi_ {abc}\psi_ {def} - 
 G^ {ab} G^ {ce} G^ {df}\psi_ {abc}\psi_ {def}+ 
 \frac{2}{3} G^ {cf} \Theta^ {ad} \Theta^ {be}\psi_ 
{abc}\psi_ {def}\nonumber\\&&\qquad\qquad\qquad\quad  - 
\frac{8}{3} G^ {bc} G^ {de} \Theta^ {ai}\psi_ {bcd} \omega_ 
{aei} + G^ {cd} \Theta^ {ai} \Theta^ {bj} \omega_ 
{acj} \omega_ {bdi} + 
 \frac{4}{3} G^ {bj} \Theta^ {ai} \Theta^ {cd} \omega_ 
{acj} \omega_ {bdi} \nonumber\\&&\qquad\qquad\qquad\quad+ 
 \frac{5}{3} G^ {cd} \Theta^ {ai} \Theta^ {bj} \omega_ 
{aci} \omega_ {bdj} - 
 \frac{8}{3} G^ {cd} \Theta^ {ai} \Theta^ {bj} \omega_ 
{abi} \omega_ {cdj} + 
 \frac{4}{3}  \bot^{ij} \Theta^ {ac} \Theta^ {bd} 
\omega_ {abi} \omega_ {cdj}  \nonumber\\&&\qquad\qquad\qquad\quad  + 
 2 G^ {bd} G^ {ce} \Theta^ {ai}\psi_ {bac} \omega_ 
{dei}- 2 G^ {bc} G^ {de} \Theta^ {ai}\psi_ {bac} 
\omega_ {dei}   - 
 \frac{2}{3} G^ {ai} G^ {ce} \Theta^ {bd}\psi_ {bac} \omega_ 
{dei}\nonumber\\&&\qquad\qquad\qquad\quad - 2 \Theta^ {ai} \Theta^ {bd} 
\Theta^ {ce}\psi_ {bac} \omega_ {dei}\bigg)+O(\alpha'^2)\bigg]\labell{final}
\eeqa
 When the transverse scalar fields are zero, only the terms in the second line are survived.  They are consistent with the couplings found in \cite{Andreev:2001xx, Wyllard:2001ye}. The last terms is exactly the one in \cite{Wyllard:2001ye}, the sign of the first term is different  because the sign of $\Omega^2$ in \cite{Wyllard:2001ye} is different from the one in \reef{DBI1}. The second term which is zero upon using the first order equation of motion, has not been considered in \cite{Wyllard:2001ye}.

\section{Discussion}

In this paper, we have proposed an action to all orders of gauge field and transverse scalars which is invariant under T-duality, is consistent with all orders of scalar couplings known in the literature and is consistent with the S-matrix element of four gauge bosons.  The action, however, is not covariant under the general coordinate transformations. On the other hand, if one imposes the condition that the action should be invariant under the general coordinate transformations and is consistent with the S-matrix elements, there are many unknown coefficients in the action even at the four field level. The general covariance and the S-matrix element constraints can not   fix the coefficients of the couplings. 

Since the   T-duality transformations on the gauge field and on the transverse scalars are linear, the T-dual Ward identity of S-matrix elements indicates that the contact terms of the S-matrix elements of an arbitrary number of   gauge bosons  must satisfy on-shell T-duality relations. That means the couplings must be consistent with an action which is invariant under the T-duality transformation. On the other hand, the contact terms of the S-matrix elements must reproduce all orders of the scalar couplings in \reef{DBI1}. That means the consistency of the couplings with all S-matrix elements indicates that the effective action may be invariant under the T-duality and at the same time  be consistent with all scalar couplings in \reef{DBI1}. If these constraint could not fix the coefficients of all couplings then one must impose other constraint, \ie  the consistency with the S-matrix element of gauge fields. Since the action \reef{final} is consistent with T-duality, is consistent with all scalar couplings in \reef{DBI1} and has no unknown coefficients, we    expect   this action   to be the effective action at order $\alpha'$ which includes all orders of gauge field.     In other words, we expect this action to produce contact terms of all S-matrix elements. However, the field redefinition can change the form of action to many other forms. For example, a non-covariant field redefinition may change the form of action to covariant form. So we expect the action \reef{final} to be unique up to field redefinitions.

We have found the   result \reef{final}  by imposing the T-duality constraint \reef{Tconstraint} on the action, \ie ignoring the T-duality constraint on total derivative terms. One may   impose the T-duality constraint on the Lagrangian, \ie ${\cal L}_{p-1}^{wT}-{\cal L}_{p-1}^t=0$. To this end, we have used the T-duality transformations  on $G$ and $\Theta$ that we have found in section 3. We have   found   the couplings in \reef{final} and \reef{total} with   $c_{13}=-2/3$ and $c_{11}=0$.  So the  T-duality constraint on the Lagrangian again   produces the couplings \reef{final} up to a total derivative term.   In other words, the couplings   \reef{final} are invariant under T-duality to all orders of massless  open string fields. 

The  B-field appears in the brane action in two different ways. Either through the different projections of  its field strength $H$ into world-volume and transverse space, or through the pull-back  of B-field    $P[B]$. The latter, however,  breaks the B-field gauge transformation unless it appears as the gauge invariant combination $F+P[B]$. So the gauge field strength in the action \reef{final} should be extended as $F\rightarrow F+P[B]$. However, this action   is not in covariant form, so its extension   to arbitrary metric $g_{ab}$ is not simply to replace $\eta\rightarrow g$ and to change ordinary derivatives to covariant derivatives.
 
To include the gravity into \reef{final}, one may first expand the action around flat metric and use   appropriate field redefinitions   to rewrite the couplings   in a covariant form, \eg the couplings in section 2  for two  and four  gauge fields. Then one may extend the flat space metric   into  arbitrary   metric   by extending  
 $P[\eta]\longrightarrow P[g]$ and extending  the covariant derivatives \reef{cov1} and \reef{cov2} to include  the derivatives of the metric. It would be interesting to find such covariant  form for the action and see if there is a compact expression for the couplings which includes all gauge and scalar fields. The replacement  $F\rightarrow F+P[B]$ in such covariant expression, then would produce many couplings between gravity and  B-field. 

The couplings between gravity and B-field should also appear   through the scalar curvature \cite{Kawai:1985xq,Ardalan:2002qt}.   It has been observed in \cite{Ardalan:2002qt} that the contact terms of the S-matrix element of two graviton vertex operators in the presence of constant background B-field can   be reproduced by the Riemann curvature contracted with  $G^{ab}$ and $\Theta^{ab}$. However, the metric in one of this matrices is flat metric. So the proposed couplings in  \cite{Ardalan:2002qt} includes all B-field but are not covariant as in the couplings in \reef{final}.    We expect, in order to find the couplings in covariant form, one should expand the non-covariant form of the couplings found in \cite{Ardalan:2002qt}   in   a series in terms of B-field, and then look for a covariant action which reproduces   the  series. The result should include the covariant form of the couplings \reef{final} in which  $F\rightarrow B$, as well as a series   involving the  couplings between B-field and the Riemann curvature.  

The couplings between gravity and the transverse scalar fields at order $\alpha'$ is given in \reef{DBI1}.
In this paper, we have extend the transverse scalar couplings to include the gauge field strength by making the couplings in flat space-time to be consistent with T-duality. The consistency  of the gravity part, for the special case of $\chi^i=0$, under linear T-duality   has been used in \cite{Garousi:2013gea} to find the dilaton and the   couplings involving $H=dB$. It would be interesting to study the consistency of the action \reef{DBI1} which includes all scalar and gravity couplings at order $\alpha'$ under full nonlinear T-duality. The T-dual theory should then include all couplings of   massless open and closed string fields at order $\alpha'$.

We have seen that the couplings in \reef{final} are consistent with the S-matrix element of four gauge bosons in flat spacetime. To verify that this action is consistent with  the higher order S-matrix elements, one may consider the S-matrix element of four gauge bosons in the presence of background constant B-field. The replacement  $F\rightarrow F+P[B]$ can produce such couplings in field theory. The disk-level S-matrix element of four gauge bosons with mixed boundary condition in string theory, however, produce the couplings of non-commutative gauge fields  \cite{Seiberg:1999vs,Garousi:1999ch}. So to produce such couplings from field theory \reef{final}, one should also use SW map  \cite{Seiberg:1999vs}  to transform the ordinary gauge fields in \reef{final} to the non-commutative variables, and then compare them with the corresponding S-matrix element in string theory. We leave the details of such calculation to the future works.

The leading higher derivative corrections to the DBI action in the superstring theory is at order $\alpha'^2$.   All transverse scalars corrections at order $\alpha'^2$ have been found in \cite{Bachas:1999um}. It would be interesting to apply the T-duality method used in section 3 to find the corrections which include all transverse scalars and gauge fields at order $\alpha'^2$.
 
{\bf Acknowledgments}:   This work is supported by Ferdowsi University of Mashhad under grant 2/39605(1394/11/04).

{\LARGE \bf Appendix}

In this appendix we find the low energy expansion of the S-matrix element of four gauge bosons in the bosonic string theory. The S-matrix element has been calculated  in \cite{Kawai:1985xq}, 
\beqa
{\cal{A}}_{1234}&\sim&\frac{\Gamma(-s)\Gamma(-t)}{\Gamma(1+u)}K'\labell{A}
\eeqa
where the Mandelstam variables are $s=- k_1\inn k_2$, $t=-  k_1\inn k_4$, $u=- k_1\inn k_3$ which satisfies $s+t+u=0$, and $K'$ is the following  kinematic factor:
\beqa
K'&=&-K+\{s[\z_1\inn k_3\z_2\inn k_3(\z_3\inn k_1\z_4\inn k_1+\z_3\inn k_2\z_4\inn k_2)+\frac{1}{3}(\z_1\inn k_2\z_2\inn k_3\z_3\inn k_1-\z_1\inn k_3\z_2\inn k_1\z_3\inn k_2)\nonumber\\&&\times(\z_4\inn k_1-\z_4\inn k_2)]+\frac{tu}{1+s}(\z_1\inn\z_2-\z_1\inn k_2\z_2\inn k_1)(\z_3\inn\z_4-\z_3\inn k_4\z_4\inn k_3)-tu\z_1\inn\z_2\z_3\inn\z_4\}\nonumber\\
&&+\{1,2,3,4\}\rightarrow \{1,3,2,4\}+\{1,3,2,4\}\rightarrow \{2,3,1,4\}
\eeqa
where $K$ is the kinematic factor of the corresponding superstring amplitude. It has four momenta and plays no role in the couplings of four gauge fields at  the order of six momenta as we will see shortly. In above amplitude $\alpha'=1/2$. Both $K$ and $K'$ are $stu$ symmetric.

The $\alpha'$-expansion of the Gamma functions is 
\beqa
\frac{\Gamma(-s )\Gamma(-t )}{\Gamma(1+u)}&=&\frac{1}{st}-\frac{\pi^2}{6}-\z(3)(s+t)-\frac{\pi^4}{360}(4s^2+st+4t^2)+\cdots\nonumber
\eeqa
The total amplitude includes all non-cyclic permutation of the external states, \ie
\beqa
{\cal{A}}={\cal{A}}_{1234}+{\cal{A}}_{1243}+{\cal{A}}_{1324}+{\cal{A}}_{1342}+{\cal{A}}_{1423}+{\cal{A}}_{1432}
\eeqa
Using the relation $s+t+u=0$, one finds that  ${\cal{A}}$ has no   mssless pole. It becomes
 \beqa
{\cal A}&\sim& -\bigg[\pi^2+ \frac{\pi^4}{24}(t^2+s^2+u^2)+\cdots\bigg]K' 
\eeqa
It is clear now that the superstring kinematic factor $K$ in $K'$ has no contribution at six momenta in the total amplitude. It has contribution at four, eight, and higher momenta. The four momenta terms are reproduced by the DBI action \reef{DBI} which  are proportional to $\pi^2$, and its eight momenta terms are reproduced by $\Omega^4$ terms \cite{Bachas:1999um} which are  proportional to $\pi^4$.

The $\alpha'$-expansion of the tachyon pole in $K'$, produces the following on-shell couplings  at six momenta which are proportional to $\pi^2$:
\beqa
{\cal{A}}_{6k}(A^4)&\sim&\{s[\z_1\inn k_3\z_2\inn k_3(\z_3\inn k_1\z_4\inn k_1+\z_3\inn k_2\z_4\inn k_2)+\frac{1}{3}(\z_1\inn k_2\z_2\inn k_3\z_3\inn k_1-\z_1\inn k_3\z_2\inn k_1\z_3\inn k_2)\nonumber\\&&\times(\z_4\inn k_1-\z_4\inn k_2)]- tu (\z_1\inn\z_2  \z_3\inn k_4\z_4\inn k_3+ \z_1\inn k_2\z_2\inn k_1 \z_3\inn\z_4 )-stu\z_1\inn\z_2\z_3\inn\z_4\}\nonumber\\
&&+\{1,2,3,4\}\rightarrow \{1,3,2,4\}+\{1,3,2,4\}\rightarrow \{2,3,1,4\} \labell{4A}
\eeqa
The couplings for the transverse scalars can be found from the above couplings by using the condition that the scalar polarization is in transvers space, \ie $\z_i\inn k_j=0$. So the couplings for four scalars is 
\beqa
{\cal{A}}_{6k}(\chi^4)&\sim& -stu(\z_1\inn\z_2\z_3\inn\z_4+\z_1\inn\z_3\z_2\inn\z_4+\z_1\inn\z_4\z_2\inn\z_3)\labell{4X}
\eeqa
The couplings for two gauge fields with polarizations $\z_1, \z_2$ and two scalars with polarizations $\z_3, \z_4$ is
\beqa
{\cal{A}}_{6k}(A^2\chi^2)&\sim&  - tu (  \z_1\inn k_2\z_2\inn k_1 \z_3\inn\z_4 )-stu\z_1\inn\z_2\z_3\inn\z_4   \labell{2A2X}
\eeqa
We compared the couplings \reef{4X} with four scalar couplings in \reef{DBI1} and find exact agreement. This fixes the normalization of the scatting amplitude \reef{A}. We compared the couplings \reef{2A2X} with two scalar and two gauge fields in \reef{DBI1} and \reef{DFDF} and found the unknown coeffient in \reef{DFDF} to be $\alpha=1$. 
The $\alpha'$-expansion of the tachyon pole in $K'$, produces also terms with eight, ten and higher momenta. All are proportional to  $\pi^2$. The terms with eight momenta may be reproduced by $D\Omega D\Omega$ and its T-dual completion. Such terms are absent in the superstring theory.
\bibliographystyle{/Users/Nick/utphys} 
\bibliographystyle{utphys} \bibliography{hyperrefs-final}
\providecommand{\href}[2]{#2}\begingroup\raggedright
\endgroup

\end{document}